\documentclass[useAMS,usenatbib]{mn2e}
\usepackage{graphicx,natbib}

\def\mV{\mbox{$M_{V}$}}

\def\ms{\mbox{$\rm M_\odot$}}

\def\x2{\mbox{$R_{rms}$}}
\def\nR{\mbox{$N_{run}$}}
\def\P{\mbox{$\vec{P}$}}
\def\Po{\mbox{$\vec{P}_o$}}
\def\Pm{\mbox{$\vec{P}_{min}$}}

\def\kS{\mbox{$k_s$}}
\def\tA{\mbox{$\overline t_{gc}$}}
\def\t0{\mbox{$t_o$}}
\def\fm{\mbox{$\mathcal{M}$}}
\def\mc{\mbox{$\mathcal{M}_c$}}
\def\mto{\mbox{$\mathcal{M}_{to}$}}

\title[The Galactic globular cluster MF]{From light to mass: accessing 
the initial and present-day Galactic globular cluster mass functions}

\author[C. Bonatto and E. Bica]{C. Bonatto$^1$ and E. Bica$^1$\\
$^1$ Departamento de Astronomia, Universidade Federal do Rio Grande
do Sul, Av. Bento Gon\c{c}alves 9500\\
Porto Alegre 91501-970, RS, Brazil}

\begin{document}

\pagerange{\pageref{firstpage}--\pageref{lastpage}}

\maketitle

\label{firstpage}

\begin{abstract}
The initial and present-day mass functions (ICMF and PDMF, respectively) of the Galactic 
globular clusters (GCs) are constructed based on their observed luminosities, the stellar
evolution and dynamical mass-loss processes, and the mass-to-light ratio (MLR). Under these 
conditions, a Schechter-like ICMF is evolved for approximately a Hubble time and converted 
into the luminosity function (LF), which requires finding the values of 5 free parameters: 
the mean GC age (\tA), the dissolution timescale of a $10^5\,\ms$ cluster ($\tau_5$), the 
exponential truncation mass (\mc) and 2 MLR parametrising constants. This is achieved by 
minimising the residuals between the evolved and observed LFs, with the minimum residuals 
and realistic parameters obtained with MLRs that increase with luminosity (or mass). 
The optimum PMDFs indicate a total stellar mass of $\sim4\times10^7$\,\ms\ still bound 
to GCs, representing $\sim15\%$ of the mass in clusters at the beginning of the gas-free 
evolution. The corresponding ICMFs resemble the scale-free MFs of young clusters and molecular 
clouds observed in the local Universe, while the PDMFs follow closely a lognormal distribution 
with a turnover at $\mto\sim7\times10^4$\,\ms. For most of the GC mass range, we find an MLR
lower than usually adopted, which explains the somewhat low \mto. Our results confirm that 
the MLR increases with cluster mass (or luminosity), and suggest that GCs and young clusters 
share a common origin in terms of physical processes related to formation.
\end{abstract}


\section{Introduction}
\label{Intro}

Globular clusters (GCs) are the living fossils of an epoch dating back to the time of  
Milky Way's formation and thus, they probably share the same physical conditions and 
processes that originated our Galaxy. Since formation, GCs are continuously affected 
by several sources of mass loss but, because of a large initial mass and/or an orbit 
that takes them most of the time away from the Galactic centre and/or disk, they are, 
in general, characterised by a long longevity. This suggests that some primordial 
information may have been preserved, if not at the individual scale, at least with 
respect to the collective properties shared by the surviving GC population.

In this context, an interesting issue is the origin of the Milky Way's (old) GC population 
and its relation to the young massive clusters that form in starbursts and galaxy mergers
of the local Universe. If both old and young cluster populations share a similar,
starburst origin, the mass distribution of their members should also be characterised by 
similar mass functions (MFs), something that is ruled out by observations in several local 
galaxies. For instance, when expressed in terms of the number of GCs per unit logarithmic 
mass-interval ($dN/d\log\mathcal{M}$), the MFs of young clusters in several galaxies follow 
monotonically a power law of slope $\approx-1$, while the present-day GC MF (PDMF) presents
a turnover for masses lower than $\mto\sim10^5\,\ms$ (\citealt{MF08} and references therein). 
The low-mass drop in the lognormal PDMF has been interpreted as mainly due to the large-scale 
dissolution of low-mass GCs after several Gyrs of tidal interactions with the Galaxy and 
internal dynamical processes (e.g. \citealt{MF08}; \citealt{Kruij08}; \citealt{KPZ09}).

Consequently, since mass loss continuously changes the MF shape - especially in the low-mass 
range, a direct comparison of the several Gyr-old PDMF with the MF of young clusters should 
not be used to draw conclusions on a physical relation between both populations. On the other 
hand, this raises another, but related question of whether or not the initial MFs (ICMFs) of 
globular and young clusters are similar, since theoretical evidence suggest that cluster 
formation leaves distinct imprints in the ICMFs (\citealt{ElmFal96}). Thus, if both cluster 
populations have the same origin, this should be reflected in the respective ICMFs.

Given the above, the reconstruction of Milky Way's GC ICMF requires having access not only to 
the (as complete as possible) PDMF, but also to the mass-loss physics. As for the latter, recent 
numerical simulations and semi-analytical models have produced robust relations for the mass 
evolution of star clusters located in different environments and losing mass through stellar 
evolution and tidal interactions with Galactic substructures and giant molecular clouds (e.g. 
\citealt{Lamers05}; \citealt{Lamers10}). 

Over the years, the GCMF time evolution has been investigated with a wide variety of approaches. 
For instance, \citet{Vesperini98} takes the effects of stellar evolution, two-body relaxation, 
disc shocking, dynamical friction, and the location in the Galaxy into account to follow the 
evolution of individual GCs. The main findings are: the large-scale disruption of GCs in the 
central parts of the Galaxy leads to a flattening in their spatial distribution; if the initial 
GCMF shape is lognormal, it is preserved in most cases; and in the case of an initial power-law 
GCMF, evolutionary processes tend to change it into lognormal. \citet{Baumgardt98} considers GC 
systems formed with a power-law MF and follows the orbits of individual clusters. Taking several 
GC destruction mechanisms into account, he finds luminosity distributions that depend on Galactocentric 
distance. \citet{VZKA03} study the dynamical evolution of M\,87 GCs with numerical simulations that
cover a range of initial conditions for the GCMF and for the spatial and velocity distribution
of the GCs. The simulations include the effects of two-body relaxation, dynamical friction, and 
stellar evolution-related mass loss. They confirm that an initial power-law GCMF can be changed  
into bell-shaped through dynamical processes, and show that the inner flattening observed in the 
spatial distribution of M\,87 GCs can result from the dynamical evolution on an initially steep 
density profile. \citet{PG07} show that an approximately Gaussian MF is naturally generated from 
a power-law mass distribution (but with a lower-mass limit) of protoglobular clouds by expulsion 
from the protocluster of star-forming gas due to supernova activity. \citet{SKT08} targets the 
MF and radial distribution evolution of the Galactic GCs by means of a Fokker-Planck model that 
considers dynamical friction, disc/bulge shocking and eccentric GC orbits. Their best-fit models 
predict the initial GC mass as $M_{GC}^0\sim(1.5-1.8)\times10^8\,\ms$. 

At some point, most of the previous works compare models with the {\em observed} GC PDMF.
However, especially because of the large distances 
involved, a direct determination of the stellar mass for most GCs is not available, which means 
that the observational PDMF cannot be taken as an ideal starting point for this analysis. Instead, 
the most direct and robust observable available for the present-day Milky Way's GC population 
as a whole is the luminosity function (LF), which is usually expressed in terms of the absolute 
$V$ magnitude. Obviously, the use of luminosity requires a mass-to-light relation 
(MLR) to convert LFs into MFs.

Recent observational and theoretical findings have convincingly shown that the MLR is not 
constant with respect to cluster mass. Instead, the dynamical evolution of clusters - and 
the preferential loss of low-mass stars - also affects the MLR. Specifically, for a given 
age and metallicity, the MLR is expected to increase with cluster mass (or luminosity; e.g. 
\citealt{Kruij08}; \citealt{KPZ09}). Interestingly, the MLR dependence on luminosity 
(or mass) is already suggested by the dynamical cluster mass and MLR  estimates of 
\citet{PM93}.

Our goal with the present paper is to reconstruct both the ICMF and PDMF of the Galactic GCs, 
starting with their observed LF and taking the mass loss into account. In this process, 
different analytical forms of the MLR will be tested to search for constraints.

The present paper is organised as follows. In Sect.~\ref{GCLF} we build the observed and
theoretical LFs of the Galactic GCs, convert the latter into the PDMF, and discuss the adopted 
MLRs. In Sect.~\ref{ASA} we briefly describe the parameter-search method. In Sect.~\ref{results} 
we discuss the results. Concluding remarks are given in Sect.~\ref{Conclu}.

\section{The theoretical present-day LF}
\label{GCLF}

The first step in our analysis is to build the observed (or present-day) LF ($\phi(M_V)\equiv
\,d\,N/d\,M_V$, i.e., number of GCs per unit absolute magnitude interval) of the Galactic GCs. 
The data were taken from the 2010 revision of the catalogue compiled by \citet{Harris96}, 
which contains 156 GCs with {\em bona-fide} absolute magnitudes in the V band (\mV). By 
definition, the integral of $\phi(M_V)$, over the full $M_V$ range is the input number of 
GCs. As a caveat, we note that several works indicate that not all Galactic GCs share
a common history (e.g. \citealt{BBBO06} and references therein), with some having been
accreted from dwarf galaxies (e.g. \citealt{HR00}; \citealt{PJMNT00}) and others associated 
with the Galactic thick disc (e.g. \citealt{Zinn85}). 

\begin{figure}
\resizebox{\hsize}{!}{\includegraphics{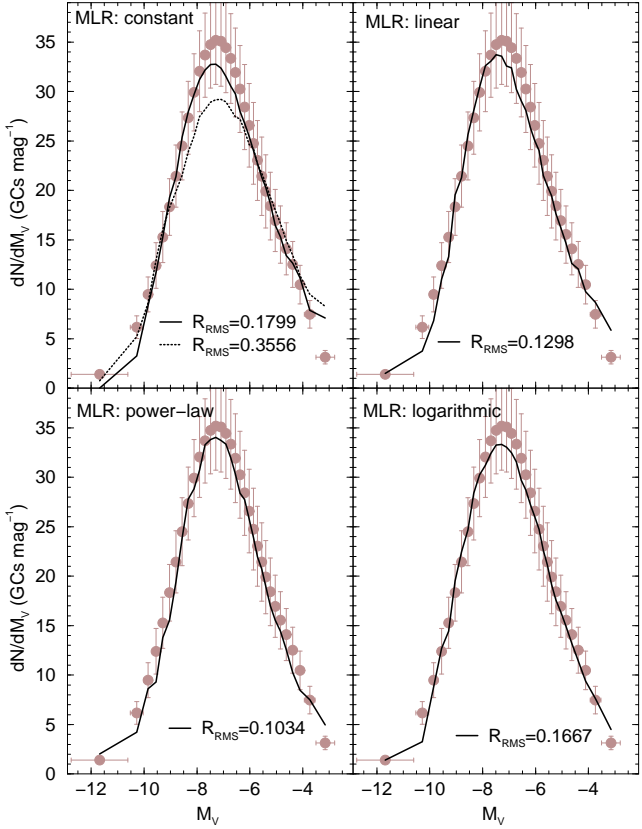}}
\caption[]{The observed (light-shaded symbols) and best-fit LFs (solid line) produced with 
alternative MLRs. The respective \x2\ are shown. A constant MLR with the mean value of 
$\fm/L_V=2$ (dotted line) taken from the literature produces a relatively poor fit with 
$\x2\approx0.36$ (top-left panel). }
\label{fig1}
\end{figure}

Following previous work (e.g. \citealt{PM2011}), uncertainties in \mV\ are incorporated in the 
LF by assuming a normal (Gaussian) distribution of errors. Specifically, if the magnitude and 
standard deviation of a GC (usually, the mean value over a series of independent measurements) 
are given by $\bar M_V\pm\sigma$, the probability of finding it at a specific value $\mV$ is 
given by $P(\mV)=\frac{1}{\sqrt{2\pi}\sigma}\,e^{{-\frac{1}{2}}\left(\frac{M_V-\bar M_V}{\sigma}
\right)^2}$. Next, we build a grid of bins covering the full \mV\ range in \citet{Harris96} and, 
for each bin, we sum the \mV\ density (or probability) for all GCs (i.e., the difference of the 
error function computed at the bin borders). However, since uncertainties are not given in 
\citet{Harris96}, for simplicity we adopt relative errors that depend on the apparent 
magnitude ($m_V$) as $\epsilon = 0.1\,e^{(m_V/27)}$. Thus, relatively bright ($m_V\la10$) GCs 
have $\epsilon\la0.15$, while fainter ones ($m_V\ga20$) have $\epsilon\ga0.2$. Finally, we assume 
that the absolute error in $M_V$ is simply given by $\sigma=\epsilon\times\left|M_V\right|$. 
As shown in \citet{PM2011}, the end result is a distribution function significantly smoother than 
a classical histogram. Indeed, the observed LF (Fig.~\ref{fig1}) is approximately gaussian-shaped 
with a pronounced peak around $M_V\approx-7$ and $\approx4$\,mags of full width at half maximum.

As usual with respect to distribution functions, the observed $\phi(M_V)$ can be 
linked to the PDMF, $\xi(\fm)\equiv\,d\,N/d\,\fm$, by 

\begin{equation}
\phi(M_V) = \xi(\fm)\left|\frac{d\,\fm}{d\,M_V}\right|.
\label{actualpm}
\end{equation}
However, except for a few cases, the stellar mass of most GCs has not been directly 
measured, and so, the actual shape of $\xi(\fm)$ remains uncertain. On the other 
hand, $\xi(\fm)$ can be inferred if one knows the initial\footnote{In the present context, 
initial refers to the onset of the gas-free evolution. Hereafter, we refer to the beginning 
of this period as \t0.} MF, the GCs age, and the mass-loss physics. Formally, if $\xi_o(\fm_o)$ 
is the MF at \t0 - and assuming that GCs are not formed subsequently, the 
MF at a later time $t$ is given by

\begin{equation}
\xi(\fm,t)=\xi_o(\fm_o)\left|\frac{\partial\fm_o}{\partial\fm}\right|,
\label{eqPartial}
\end{equation}
where $\fm = \fm(\fm_o,t)$, and $\fm_o$ represents the individual GC masses at \t0. 

Several works (e.g. \citealt{PMG10}; \citealt{Larsen09}) suggest that star clusters 
form with a mass distribution that follows the truncated power-law of \citet{Schechter76} 

\begin{equation}
\xi_o(\fm_o) = \kS\fm_o^{-2}e^{-\left(\fm_o/\fm_c\right)}
\label{schechter}
\end{equation}
over the mass range $[\fm_{inf}$ -- $\fm_{sup}]$, where $\fm_c$ is the exponential truncation 
mass, and \kS\ is a normalisation constant. We note that, because of the continuous mass 
loss (see below), GCs that formed with a relatively low mass may no longer be detected, and 
can be considered as dissolved. Here, we represent this dissolution limit by the lower 
{\em observable} mass value $\fm_i$.

The earliest cluster phase is dominated by effects related to the impulsive removal
of the parental, intra-cluster gas by supernovae and massive-star winds, which rapidly 
damps cluster formation and changes the ratio between residual gravitational potential 
and stellar velocity dispersion, thus leading to the first bout of dynamical evolution 
(e.g. \citealt{LMD84}; \citealt{Goodwin97}; \citealt{GB01}; \citealt{BK07}). Another 
consequence of these early effects on the ICMF may be a depletion in the number of GCs
towards the low-mass range (e.g. \citealt{KB02}; \citealt{BKP08}; \citealt{PGKB08}). 
Clusters can be considered to be essentially gas free after the first $\t0\sim10^7$\,yrs,
but they keep losing mass mainly through stellar evolution and externally-driven tidal 
effects. After several Gyrs, these processes may reduce the mass to a fraction of the
initial value. Following 
\citet{Lamers05}, we express the mass remaining (due to stellar evolution and tidal effects) 
in a given GC (with initial mass $\fm_o$) at time $t$ by

\begin{equation}
\fm(t) = \fm_o\left[\mu_{se}^\gamma - \frac{\gamma\,t}{\tau_5}\left(\frac{\fm_o}{10^5\ms}\right)
^{-\gamma}\right]^{1/\gamma},
\label{massT}
\end{equation}
where $\mu_{se}$ is the stellar evolution mass loss, $\tau_5$ is the dissolution timescale of 
a $10^5\,\ms$ star cluster, and $\gamma$ is the power-law exponent that sets the dependence of 
the cluster disruption time on mass. We note that $\mu_{se}$ is a function of time, but 
since we are dealing with GCs with ages between 9-13\,Gyr, we simplify this point by taking here 
the asymptotic value $\mu_{se}\approx0.7$ (\citealt{Lamers05}). Useful Milky Way approximations 
for the other parameters are $\gamma\approx0.7$ (\citealt{Lamers10}), and $\tau_5\approx10^4$\,Myr
(\citealt{KC2011} and references therein). However, $\tau_5$ is expected to present a significant 
dispersion around its quoted value to account for different GC environments and orbital conditions 
(\citealt{KC2011}). In this sense, we consider $\tau_5$ to be a free parameter allowed to vary
within a given range (Sect.~\ref{ASA}). Thus, inverting Eq.~\ref{massT} to have $\fm_o=\fm_o(\fm(t))$ 
and coupling Eqs.~\ref{eqPartial} and \ref{schechter}, the PDMF can be expressed as

\begin{equation}
\xi(\fm,t) = \kS \mu_{se}
\fm^{\gamma-1} \left(\fm^\gamma + \beta\,t\right)^{-(1+1/\gamma)}
e^{-\frac{\left(\mathcal{M}^\gamma + \beta\,t\right)^{1/\gamma}}{\mu_{se}\mathcal{M}_c}},
\label{pmdf}
\end{equation}
where $\beta\equiv(\gamma10^{5\gamma})/\tau_5$.

Formally, Eq.~\ref{pmdf} applies to an individual cluster having evolved over a time $t$
after \t0. However, the underlying idea of our approach is to consider not the individual, but
the collective - or average - evolution of the full sample of the Galactic GCs until now. This
adds another component to our approach because, instead of a single formation event, the Milky 
Way GCs present some dispersion in age (e.g. \citealt{GCage}). In this sense, $t$ should be 
taken as the mean - or at least, the representative - age (\tA) of the GC population. So, hereafter 
we consider $t=\tA$. The constant \kS\ can be computed by noting that the integral of Eq.~\ref{pmdf} 
over the minimum observable GC mass, $\fm_i$, and the maximum remaining mass at time $\tA$, 
$\fm_s = \fm_{sup}(\tA)$, corresponds to the present-day number of GCs ($N_{GC}$) 

\begin{equation}
N_{GC} = \int_{\mathcal{M}_i}^{\mathcal{M}_s}\xi(\fm,\tA)\,d\,\fm.
\label{ngc}
\end{equation}
 
Changing variables in the integral, Eq.~\ref{ngc} can be reduced to $$ N_{GC} = \frac{\kS}{\fm_c} 
\int_{\tau_i}^{\tau_s}\left[\tau\ln(\tau)\right]^{-2}d\,\tau =\frac{\kS}{\fm_c}\left[E1\,(\ln(\tau)) 
- \frac{1}{\tau\ln(\tau)}\right]_{\tau_i}^{\tau_s},$$
where $E1$ is the exponential integral, 
$$\ln(\tau_i) = \left(\fm_i^\gamma + \beta\,\tA\right)^{1/\gamma}/\left(\mu_{se}\mathcal{M}_c\right),$$
and
$$\ln(\tau_s) = \left(\fm_s^\gamma + \beta\,\tA\right)^{1/\gamma}/\left(\mu_{se}\mathcal{M}_c\right).$$

Summarising, the predicted PDMF (Eq.~\ref{pmdf}) can be inserted into 
Eq.~\ref{actualpm} to build the theoretical LF $\widetilde\phi(M_V)$ and, with the usual 
relation between luminosity ($L_V$) and absolute magnitude $L_V=10^{-0.4(M_V-M_V^\odot)}$,
with $M_V^\odot=4.83$, the theoretical LF is expressed as 

\begin{equation}
\widetilde\phi(M_V) = 0.4\,\xi(\fm,\tA)\,\frac{d\,\fm}{d\,\log{L_V}}.
\label{pdlf}
\end{equation}

Subsequently, $\widetilde\phi(M_V)$ can be compared with the observed $\phi(M_V)$ to search for the 
parameters that lead to the best match between both. Note that the present-day GC masses in 
Eq.~\ref{pdlf} should be expressed as a function of $M_V$ (through $L_V$), which can be 
done by means of a mass-to-light relation ($MLR\equiv \fm/L_V$). 

\subsection{The mass-to-light ratio}
\label{MLR}

According to Sect.~\ref{Intro} and Eq.~\ref{pdlf}, it is clear that the dependence of MLR 
on luminosity (or mass) is central to the task of building MFs from the observed LF. In 
this sense, we consider here the following comprehensive range of MLR shapes, {\em (i)}
constant: $\fm/L_V = a$; {\em (ii)} linearly-increasing with luminosity: $\fm/L_V = a + b\,L_V$;
{\em (iii)} power-law: $\fm/L_V = a + b\,L_V^n$; and {\em (iv)} logarithmic: $\fm/L_V = a + 
b\,\log(L_V)$, where $a$ and $b$ are constants. Note that for practical reasons (i.e., the 
need of an analytical expression to convert $M_V$ into $\fm$ in Eq.~\ref{pdlf}), we express 
the MLR as a function of $L_V$ and not mass. 

Obviously, once the MLR constants $a$ and $b$ have been assigned values, the minimum and 
maximum present-day GC masses ($\fm_i$ and $\fm_s$) are naturally determined from the 
respective faint and bright boundaries of the observed LF. And, for a given mean GC age 
(\tA), a similar reasoning leads to the maximum GC mass ($\fm_{sup}$) at \t0\ through
inversion of Eq.~\ref{massT}. 


\section{Searching for the optimum parameters}
\label{ASA}

The approach summarised by Eq.~\ref{pdlf} contains some poorly-known (and unknown) parameters, 
such as the mean cluster age ($\tA$), the dissolution timescale ($t_5$), the truncation mass 
($\fm_c$), and the MLR constants $a$ and $b$. Our approach considers these five parameters as 
free. However, we note that some dependence among these parameters should be expected to 
occur, especially for very wide search ranges. For instance, a short \tA\ would result in a 
low $\tau_5$ to hasten dissolution over a shorter age range. Similarly, a lower MLR would shift 
the MF turnover to a lower cluster mass, which could result from a longer dissolution timescale 
(i.e. less dynamical evolution) for a given mean cluster age. We minimise this effect by adopting 
a reasonable GC age range (see below).

Thus, the task now is to search for values for the set of free parameters 
$\P\equiv\left\{a, b, \tA, t_5, \fm_c\right\}$ that produces the best match 
between the observed and theoretical LFs, $\phi(M_V)$ and $\tilde\phi(M_V)$, respectively. 
This is achieved by minimising the root mean squared residual (\x2) between both 
functions. By construction, \x2\ is a function of the free parameters, i.e., $\x2=\x2(\P)$, 
which we express as

\begin{equation}
\label{eqRMS}
\x2=\sqrt{\frac{1}{N_{GC}}\sum_{j=1}^{n_b}\frac{\left[\tilde\phi(M^j_V)-\phi(M^j_V)\right]^2}
{\tilde\phi(M^j_V)+\phi(M^j_V)}},
\end{equation}
where the sum occurs over the $n_b$ non-zero magnitude bins of the observed LF. The 
normalisation by the theoretical$+$observed density of GCs in each bin gives a higher weight 
to the more populated bins\footnote{In Poisson statistics, where the uncertainty ($\sigma_\phi$) 
of a signal $\phi$ is $\sigma_\phi=\sqrt{\phi}$, our definition of \x2\ turns out being equivalent 
to the usual $\chi^2$.}. Finally, the squared sum is divided by the total number of observed GCs, 
which makes \x2\ dimensionless and preserves the number statistics when comparing LFs built with 
unequal GC populations. 

\begin{table*}
\caption[]{Present-day and initial MF parameters found with alternative MLRs}
\label{tab1}
\renewcommand{\tabcolsep}{1.25mm}
\renewcommand{\arraystretch}{1.25}
\begin{tabular}{ccccccccccccc}
\hline\hline
     &&\multicolumn{5}{c}{Free Parameters}&\multicolumn{5}{c}{Derived Parameters}\\
\cline{3-7}\cline{9-13}\\
Rank&$\x2$& \tA &    $a$     & $b$ & $\fm_c$ & $\tau_5$&& $\fm_i$   & $\fm_{sup}$       & $N_{GC}^0$ &  $M_{GC}^0$   & $M_{GC}^{now}$ \\
    & &(Gyr)&            &     & (\ms) &   (Gyr) &&($10^3\,\ms$)&$(10^7\,\ms)$    & ($10^4$)   & ($10^8\,\ms$) &  ($10^6\,\ms$)   \\
(1)&(2)&(3)&(4)&(5)&(6)&(7)&&(8)&(9)&(10)&(11)&(12)\\

\hline
\multicolumn{13}{c}{Constant: $\fm/L_V=a$}\\
&      &              &              &                            & $(\times10^5)$\\
\hline
Abs.Min.                &0.1799&  10.7        &  0.59         & --- &  6.0        &  5.8        &&  0.7        &  0.5        & 9.9         &  4.4        &  9.2 \\
$\rm\overline{1\%~Low.}$&0.1875& $11.4\pm1.0$ & $0.49\pm0.19$ & --- & $5.1\pm2.0$ & $8.1\pm4.3$ && $0.7\pm0.3$ & $0.4\pm0.2$ & $8.4\pm1.7$ & $3.7\pm1.4$ & $8.3\pm3.2$ \\
$\rm\overline{Overall}$ &0.2041& $11.3\pm1.0$ & $0.50\pm0.23$ & --- & $5.5\pm2.7$ & $8.7\pm5.3$ && $0.8\pm0.4$ & $0.4\pm0.3$ & $7.5\pm2.0$ & $3.4\pm1.5$ & $8.4\pm3.9$  \\

\hline
\multicolumn{13}{c}{Logarithmic: $\fm/L_V=a+b\,\log(L_V)$}\\
&      &              &$(\times10^{-1})$             & &$(\times10^6)$\\
\hline
Abs.Min.                &0.1667&  10.3        &  0.32         &  0.19         &  1.6        &  5.0        &&  0.7        &  1.2        &  7.3        &  3.8        &  17 \\
$\rm\overline{1\%~Low.}$&0.1772& $11.1\pm1.1$ & $0.19\pm0.21$ & $0.15\pm0.05$ & $1.3\pm0.5$ & $7.8\pm4.5$ && $0.8\pm0.3$ & $0.7\pm0.4$ & $5.8\pm1.2$ & $2.9\pm1.0$ & $14\pm5$ \\
$\rm\overline{Overall}$ &0.1936& $11.4\pm1.0$ & $0.65\pm0.66$ & $0.16\pm0.05$ & $1.5\pm0.6$ & $6.9\pm2.7$ && $0.9\pm0.3$ & $0.8\pm0.5$ & $5.3\pm1.1$ & $3.1\pm1.0$ & $16\pm5$  \\
          
\hline
\multicolumn{13}{c}{Linear: $\fm/L_V=a+b\,L_V$}\\
&      &              &              &$(\times10^{-5})$&$(\times10^{10})$\\
\hline
Abs.Min.                &0.1298&  12.6        &  0.43         &  0.20         &  1.6        &  5.4        &&  0.6        &  26       &  1.9        &  1.2        &  21 \\
$\rm\overline{1\%~Low.}$&0.1345& $11.4\pm1.1$ & $0.91\pm0.47$ & $0.45\pm0.23$ & $5.2\pm2.8$ & $9.7\pm6.1$ && $1.4\pm0.8$ & $13\pm15$ & $1.7\pm0.6$ & $2.7\pm1.4$ & $41\pm22$ \\
$\rm\overline{Overall}$ &0.1543& $11.3\pm1.1$ & $0.93\pm0.44$ & $0.46\pm0.22$ & $5.0\pm2.8$ & $9.1\pm5.7$ && $1.5\pm0.8$ & $18\pm22$ & $1.7\pm0.3$ & $1.7\pm1.3$ & $42\pm20$  \\

\hline
\multicolumn{13}{c}{Power-law: $\fm/L_V=a+b\,L_V^{0.67}$}\\
&      &              &            &$(\times10^{-3})$&$(\times10^{10})$\\
\hline
Abs.Min.                &0.1034&  10.5        &  0.83         &  0.36         &  1.6        &  5.8        &&  1.8        &  5.3        &  1.7        &  3.1        &  44 \\
$\rm\overline{1\%~Low.}$&0.1213& $11.3\pm0.9$ & $0.72\pm0.33$ & $0.29\pm0.14$ & $5.1\pm2.8$ & $8.8\pm4.8$ && $1.2\pm0.6$ & $7.0\pm6.7$ & $2.2\pm0.4$ & $2.7\pm1.3$ & $37\pm17$ \\
$\rm\overline{Overall}$ &0.1395& $11.3\pm0.8$ & $0.71\pm0.34$ & $0.28\pm0.13$ & $5.0\pm2.8$ & $9.0\pm5.5$ && $1.2\pm0.6$ & $6.8\pm6.8$ & $2.1\pm0.4$ & $2.6\pm1.2$ & $36\pm17$  \\
          
\hline
\end{tabular}
\begin{list}{Table Notes.}
\item Cols.~(1) and (2): Rank and corresponding \x2\ value; Col.~(3): mean GC age; Cols.~(4)
and (5): MLR constants; Col.~(6) exponential truncation mass; Col.~(7): dissolution timescale 
of a $10^5\,\ms$ star cluster; Col.~(8): lowest present-day (observed) GC mass; Col.~(9): upper 
cluster mass at \t0; Cols.~(10) and (11): number and total mass of GCs at \t0\ (with
$\fm_{inf}\ga10^2\,\ms$); Col.~(12): present-day total GC mass.
\end{list}
\end{table*}

Among the several minimisation methods available in the literature, the adaptive simulated 
annealing (ASA) is adequate to our purposes, because it is relatively time efficient, robust 
and capable to distinguish between different local minima (e.g. \citealt{ASA}). The first
step in the minimisation process is to define individual search ranges (and variation steps) 
for each free parameter. Next, initial values are randomly selected for all free parameters
(i.e., the initial {\em point} \Po\ in the \x2\ hyper-surface) and the starting value of \x2\ 
is computed. Then ASA takes a step by changing the initial parameters ($\Po\rightarrow\P$), 
and a new value of \x2\ is evaluated. Specifically, this implies that a new LF is built with 
the changed parameters. By definition, any step that decreases \x2\ (downhill) is accepted, 
with the process repeating from this new point. However, uphill steps may also be taken, with 
the decision made by the Metropolis (\citealt{Metropolis53}) criterion, which has the advantage 
of enabling ASA to escape from local minima. The variation steps become smaller as the minimisation 
is successful and ASA approaches the global minimum ($\P\rightarrow\Pm$).

The output of a single ASA run is the set of optimum values \Pm\ (expected to correspond 
to the global minimum) and the respective \x2, but with no reference to uncertainties. 
However, given the errors associated to the observed LF (Fig.~\ref{fig1}), uncertainties 
to the derived parameters are strongly required. Thus, we run ASA several times (\nR) 
allowing for different starting points (\Po) and taking the uncertainty of each $M_V$ bin 
in the observed LF (Fig.~\ref{fig1}) into account. In the end, we compute the weighted 
mean of the parameters over a range of runs, using the \x2\ of each run as weight
($w=1/\x2$). As a compromise between statistical significance and computational time, we
adopt here $\nR=6000$.

The search ranges adopted here try to emulate the physical conditions prevailing in the 
Milky Way. In this sense, the mean GC age runs through $9\le\tA(Gyr)\le13$; the exponential 
truncation mass within $10^4\le\fm_c(\ms)\le10^{11}$; and the dissolution timescale within 
$5\le\tau_5(Gyr)\le30$ (to account for different environments and orbital conditions). 
Initially, the MLR constants $a$ and $b$ were allowed to vary within $0$ and $10$ but, 
after a few runs, their ranges were fine-tuned to more constrained ranges.

\section{Results and discussion}
\label{results}

The approach described in Sect.~\ref{ASA} was applied to the four MLR modes discussed 
in Sect.~\ref{MLR}, and the numerical results of the 6000 independent ASA runs are 
summarised in Table~\ref{tab1}. For conciseness, we show only the parameters belonging 
to the run with the absolute minimum \x2, together with those averaged over all the 
6000 runs, and those corresponding to the average of 1\% of the runs having the lowest 
\x2. The LFs corresponding to the absolute minimum \x2\ are compared to the observed LF 
in Fig.~\ref{fig1}. Given the relative proximity among the minimum and average parameters 
(Table~\ref{tab1}), the respective LFs of each MLR should also be similar.

\begin{figure}
\resizebox{\hsize}{!}{\includegraphics{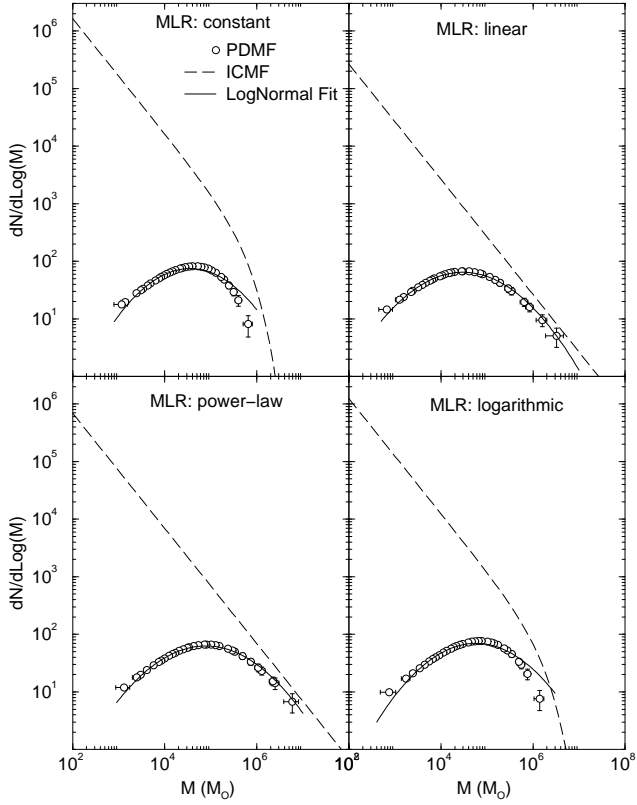}}
\caption[]{Present-day (empty circles) and initial (dashed-line) MFs that result from 
different MLRs. The PDMFs (expressed as $d\,N/d\,\log(\mathcal{M})$) are fitted with 
lognormal curves (thin solid line). }
\label{fig2}
\end{figure}

Qualitatively, the adopted MLRs produce similarly evolved LFs (Fig.~\ref{fig1}), but with 
different individual and integrated parameters at the output. As expected, the mean age, 
in all cases, is consistent with the age of the Galaxy and the age spread of the Galactic 
GCs (\citealt{MuGne2010}), ranging within 10 and 12.6\,Gyr. Also, the best-fit dissolution 
timescale falls in the range $5\le\tau_5(Gyr)\le10$, a spread that is consistent with the 
different environments where the Galactic GCs dwell. Below we discuss the results for the 
MLR modes in a decreasing order of \x2. 

Before doing so, it is useful to gather some integrated Galactic GC values to be used for
comparison with our results. For instance, based on the uniform ratio $\fm/L_V=2$,
\citet{MvdB05} estimate that the current total mass of halo GCs as $2\times10^7\,\ms$. 
However, this may be a lower limit to the actual value of $M_{GC}^{now}$, since halo
GCs correspond to a sub-population of the Galactic GCs (e.g. \citealt{MvdB05}) and SSP 
models (e.g. \citealt{BruChar03}; \citealt{AA03}) usually predict the higher ratio 
$\fm/L_V=2-4$. Also, the initial total mass in GCs ($M_{GC}^0$) can be compared with 
estimates of the current stellar halo mass of $\mathcal{M}_h\approx1\times10^9\,\ms$ 
(e.g. \citealt{FKBH02}).

{\tt Constant $\left[\fm/L_V=a\right]$:} With $\x2\sim0.18-0.2$, the best fit was obtained 
with $\fm/L_V\approx0.6$. For such MLR, the lowest present-day GC mass turns out being 
somewhat low for a GC, $\fm_i\sim700\,\ms$; the exponential truncation mass is
$\fm_c\approx6\times10^5\,\ms$. Reflecting the low statistical significance of the 
constant MLR, the dispersion around the mean for most parameters is very large. When 
combined, the derived parameters imply an exceedingly-low (see above) total mass for the 
present-day GC population, $M_{GC}^{now}\la1\times10^7\,\ms$. Assuming 100\,\ms\ as
the lower-limit for cluster formation, the total stellar mass stored in clusters at \t0\ 
would be $M_{GC}^{0}\sim4\times10^8\,\ms$. Thus, the fraction of the initial stellar mass 
still bound in GCs would be as low as $f_{bound}\approx2\%$. The higher ratio $\fm/L_V=2$ 
(the average used in previous works, e.g. \citealt{Kruij08}) leads to an evolved LF that 
fails to describe the observations, especially around the peak ($M_V\approx-7$), with the 
high value of $\x2\sim0.36$.

{\tt Logarithmic $\left[\fm/L_V=a+b\,\log(L_V)\right]$:} With \x2\ similar to the constant 
MLR, the logarithmic MLR yields $\fm_i\sim800\,\ms$, $\fm_c\sim1.5\times10^6\,\ms$,
$M_{GC}^{now}\sim1.5\times10^7\,\ms$, and $M_{GC}^{0}\sim3\times10^8\,\ms$, which implies 
the low bound mass fraction of $f_{bound}\sim5\%$. These values are similar to those
obtained with the constant MLR and, as such, may not be realistic.

{\tt Linear $\left[\fm/L_V=a+b\,L_V\right]$:} Dropping somewhat to $\x2\sim0.13-0.15$, the 
linearly increasing MLR predicts the more realistic values of $\fm_i\sim1.4\times10^3\,\ms$ 
and $M_{GC}^{now}\sim4\times10^7\,\ms$. With $M_{GC}^{0}\sim2\times10^8\,\ms$, the fraction 
of stellar mass still bound in GCs rises to $f_{bound}\sim20\%$. Interestingly, the best-fits 
require an exponential truncation mass of $\fm_c\sim10^{10}\,\ms$, about 4 orders of magnitude
higher than usually adopted (e.g. \citealt{MF08}).

\begin{figure}
\resizebox{\hsize}{!}{\includegraphics{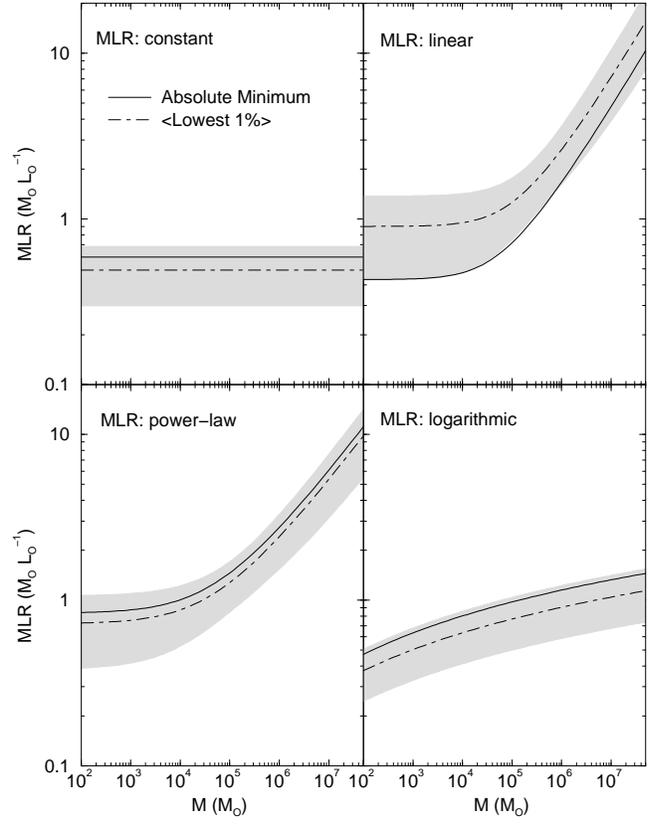}}
\caption[]{The best-fit MLRs are shown as a function of the cluster mass. For comparison,
the curves corresponding to the absolute minimum (solid line) and the 1\% runs with the
lowest \x2\ values (dot-dashed) are shown, together with the $1\sigma$ uncertainty 
domain of the latter.}
\label{fig3}
\end{figure}

{\tt Power-law $\left[\fm/L_V=a+b\,L_V^n\right]$:} To minimise the number of free 
parameters, we applied our approach for several fixed exponents $n$, with the lowest 
values of \x2\ ($\sim0.10-0.14$) obtained with $n=0.67$. The relevant best-fit parameters are 
$\fm_i\sim10^3\,\ms$, $\fm_c\sim5\times10^{10}\,\ms$, $M_{GC}^{now}\sim4\times10^7\,\ms$, 
and $M_{GC}^{0}\sim3\times10^8\,\ms$, with a bound-mass fraction of $f_{bound}\sim15\%$.

Both the linear and power-law models predict values of $M_{GC}^{now}$ that are
compatible with independent estimates. They also imply an initial total GC mass that
corresponds to $\la30\%$ of the current stellar halo mass.

Reflecting the (cluster) mass-dependent nature of the mass-loss processes, GCs having
a mass within $(1-2)\times10^5\,\ms$ at the onset of the gas-free phase would have evolved
to become the present-day least massive ($\fm_i\sim1\times10^3\,\ms$) GCs. This means that,
after a Hubble time of evolution in the Galaxy, GCs occupying the low-mass tail of the PDMF 
retain only $\sim1\%$ of their initial stellar mass. Regarding the total stellar mass still 
bound in GCs, the best-fit MLRs (power-law and linear) imply that $\sim85\%$ of the mass of 
the initial GC population would have been lost to the field. 

Finally, the best-fit MLRs are shown in Fig.~\ref{fig3} as a function of the cluster 
mass. Both the constant and logarithmic MLRs are somewhat shallow, with $\fm/L_V\la2.0$ 
over the full range of mass ($10^2\le\fm(\ms)\le10^7$). On the other hand, the
linear and power-law MLRs have $\fm/L_V\approx1.0$ for $10^2\le\fm(\ms)\la10^4$,
then rising monotonically for larger masses, reaching $\fm/L_V\approx2.5$ for 
$\fm=10^6\,\ms$, and $\fm/L_V\approx4-8$ for $\fm=10^7\,\ms$. When converted to a 
function of GC mass, the power-law MLR follows the relation 
$$\fm/L_V=(0.78\pm0.01)+\left[\frac{\mathcal{M}}{(2.17\pm0.06)\times10^5\,\ms}\right]^{0.42\pm0.01}.$$

In summary, the best results in terms of minimum \x2\ residuals and realistic GC 
parameters are obtained with MLRs that are relatively low for most of the GC mass range,
but increase with luminosity (or mass), in the 
present case, expressed as a power-law (exponent 0.67) or linearly.

\subsection{The present-day MF}
\label{PDMF}

One of the main results of our approach is to provide access to the PDMF through 
Eq.~\ref{pmdf} and the best-fit parameters of each MLR mode. 
For practical reasons, they are shown in terms of $d\,N/d\,\log(\mathcal{M})$ in 
Fig.~\ref{fig2}. They all clearly resemble a lognormal mass distribution\footnote
{Expressed as $\frac{d\,N}{d\,\log(\mathcal{M})}\propto e^{-0.5\left
(\frac{\log(\mathcal{M})-\overline{\log(\mathcal{M})}}{\sigma}\right)^2}$}, which is 
confirmed by the corresponding fit - the parameters are given in Table~\ref{tab2}. Again, 
the lognormal character applies especially to the MFs produced by the power-law and linear 
MLRs, followed by the logarithmic and constant modes. 

\begin{table}
\caption[]{Lognormal parameters of the present-day MF}
\label{tab2}
\tiny
\renewcommand{\tabcolsep}{2.1mm}
\renewcommand{\arraystretch}{1.25}
\begin{tabular}{lcccccc}
\hline\hline
MLR&  $\mto$  &$\sigma$&CC&$\overline{\mathcal M}$&$\overline{\mathcal M}/\mto$\\
        &($10^4\,\ms$)&(\ms)&  &($10^5\,\ms$)& \\
  (1) & (2) & (3) & (4) & (5) & (6) \\
\hline
Constant   &$3.5\pm0.2$&$0.80\pm0.02$&0.970&0.6&1.6\\
Logarithmic&$5.7\pm0.4$&$0.86\pm0.02$&0.989&1.0&1.8\\
Linear     &$3.1\pm0.2$&$0.90\pm0.01$&0.997&2.7&8.6\\
Power-law  &$7.8\pm0.2$&$0.91\pm0.01$&0.997&2.5&3.2\\
\hline
\end{tabular}
\begin{list}{Table Notes.}
\item Col.~(2): turnover mass; Col.~(3): dispersion in $\log(\fm)$; Col.~(4):
correlation coefficient of the lognormal fit to the PDMFs; Col.~(5):
average GC mass; Col.~(6): ratio between the mean and turnover mass.
\end{list}
\end{table}

Overall, the PDMF mass turnover predicted by our MLRs occur within
$\mto\sim(3-8)\times10^4$\,\ms\ (the upper bound corresponding to the power-law MLR), a range somewhat lower 
than the $\mto\sim(1-2)\times10^5$\,\ms\ derived in previous works (e.g. \citealt{MF08}). 
Probably, the difference occurs because for most of the GC mass range, our MLR models 
are lower than the $\fm/L_V\simeq2$ usually employed in previous studies, which naturally 
shifts \mto\ to lower values, implying longer dissolution timescales. In addition, by not 
including the effects of shocks with giant molecular clouds and spiral arms, and two-body 
evaporation and ejection\footnote{Although molecular clouds and spiral arms may not be 
relevant to most GCs, evaporation does play a role (e.g. \citealt{MF08}).}, Eq.~\ref{massT} 
probably yields a softer mass-loss rate than those previously used. We also note that, 
for masses larger than that of the turnover, the PDMFs (of the power-law and linear MLRs) 
have a slope similar (within the uncertainties) to the MFs of young clusters and molecular 
clouds observed in the Milky Way and other galaxies (\citealt{MF08}, and references therein). 
A similar high-mass slope was found for a sample of Galactic and M\,31 GCs by \citet{McLP96}.
Actually, this is a natural consequence of the large $\fm_c$, which favours the formation of 
a higher number of massive clusters with respect to a low-$\fm_c$ ICMF. Massive clusters are 
naturally long-lived and, thus, the slope of the large-mass tail of the ICMF tends to be 
preserved for longer periods of time.

We note that our best-fit MLR model predicts a maximum PDMF mass (Fig.~\ref{fig2}) that 
is compatible with current estimates for {\em Omega Cen} (NGC\,5139): $2.5\times10^6\,\ms$ 
(\citealt{vDv06}), $4\times10^6\,\ms$ (\citealt{PM93}), and $5.1\times10^6\,\ms$ (\citealt{MMDD95}). 
In particular, the latter estimate implies the ratio $\fm/L_V\approx4.1$, again consistent with 
the power-law MLR value for a GC with mass in the range $(2-5)\times10^6\,\ms$ (Fig.~\ref{fig3}). 
The power-law MLR predicts the mean present-day GC mass $\overline{\mathcal M}\approx2.5\times10^5\,\ms$, 
from which follows the scaling relation $\overline{\mathcal M}\sim3\mto$.

We also show in Fig.~\ref{fig2} the ICMFs reconstructed with the best-fit
parameters. Interestingly, while the ICMFs predicted by the constant and logarithmic
MLRs show the typical break at $\fm_c\sim10^5\,\ms$ of the Schechter profile, those of
the linear and power-law MLRs do not. Both are characterised by the large value of
$\fm_c\sim10^{10}\,\ms$ and, thus, they end up resembling simply as a scale-free,
power-law of slope $-1$ (or $-2$, when expressed as $d\,N/d\,\fm$) over the very wide
mass range $10^2$---$10^{10}$\,\ms. The latter is a typical feature of young cluster
MFs observed in nearby galaxies (e.g. \citealt{ZF99}; \citealt{PMG10}).

\section{Summary and conclusions}
\label{Conclu}

We present an approach to recover both the initial and present-day mass functions of the 
Galactic globular clusters, having as constraint the observed luminosity distribution. 
This is achieved by taking a few mass-loss processes into account and assuming a
mass-to-light ratio that is either constant or increases with luminosity linearly, as
a power law or logarithmically.

It starts with a Schechter-like ICMF, where the truncation ($\fm_c$) mass is a free parameter. 
Then, using a set of parameters that usually apply to the dynamical dissolution of Galactic 
star clusters, among these the mean GC age (\tA), the dissolution timescale of a $10^5$\,\ms\ 
cluster ($\tau_5$), and an MLR (parametrised by the constants $a$ and $b$), we compute the MF 
shape after evolving for a time \tA\ after the onset of the gas-free phase. The evolved MF is 
subsequently converted into the LF. Finally, we  search for the optimum parameters $\left(\tA, 
t_5, \fm_c, a, b\right)$ that minimise the residuals (\x2) between the evolved and observed LFs. 

The best results, both in terms of minimum values of \x2\ and realistic parameters for
the Galactic GCs, 
correspond to the power-law and linear MLRs, respectively. Specifically, because both MLRs 
increase with luminosity (or mass), their present-day MFs imply a total mass in GCs of the 
order of $M_{GC}^{now}\approx4\times10^7$\,\ms, which corresponds to a fraction of $\sim15\%$ 
of the initial stellar mass in GCs. On the other hand, the constant and logarithmic MLRs are 
exceedingly shallow for the full GC mass range, yielding unrealistically low values of $M_{GC}^{now}$. 

Because of the large truncation mass ($\fm_c\sim10^{10}$\,\ms), the Schechter-like ICMFs 
(of the linear and power-law MLRs) end up resembling the scale-free, power-law MFs of 
young clusters and molecular clouds in the Milky Way and other galaxies. In addition,
their PDMFs follow closely a lognormal distribution with a turnover mass at
$\mto\sim7\times10^4$\,\ms\ that, for larger masses, behave as a power-law with the same 
slope as the ICMFs. The low values of \mto, compared to previous works, follow from the 
relatively low MLRs (for most of the GC mass range) dealt with here.

Summarising, it is clear that the power-law is not the unique ICMF solution, the MLR, 
which serves to fine-tune the details, must increase with cluster mass (or luminosity), 
and that mass loss is the main factor responsible for shaping the ICMF into its present, 
lognormal form. Finally, our results suggest a common origin - in terms of physical 
processes - for GCs and young clusters.

\section*{Acknowledgements}
We thank an anonymous referee for relevant comments and suggestions.
We acknowledge financial support from the Brazilian Institution CNPq.




\label{lastpage}
\end{document}